\documentstyle[floats,multicol,aps,epsf,prb]{revtex}

\newcommand \beq  {\begin{equation}}
\newcommand \eeq  {\end{equation}}
\newcommand \bea {\begin{eqnarray} }
\newcommand \eea {\end{eqnarray}}

\begin{document}
\draft
\twocolumn[\hsize\textwidth\columnwidth\hsize\csname @twocolumnfalse\endcsname
\title{Scaling of magnetic fluctuations near a quantum phase transition }
\author{A. Schr\"{o}der$^{1,2}$, G. Aeppli$^{2,3}$,
E. Bucher$^{4,5}$, R. Ramazashvili$^{6}$, P. Coleman$^{6}$\\}
\address{$^{1}$ Physikalisches Institut, Universit\"{a}t Karlsruhe,D-76128 
Karlsruhe, Germany}
\address{$^{2}$ Ris{\o} National Laboratory, DK-4000 Roskilde, Denmark}
\address{$^{3}$ NEC, 4 Independence Way, Princeton, NJ 08540, U.S.A.}
\address{$^{4}$ Universit\"{a}t Konstanz, Konstanz, Germany}
\address{$^{5}$ Bell Laboratories, Lucent Technologies, Murray Hill, NJ 079
74, U.S.A.}
\address{$^{6}$ Serin Laboratory, Rutgers University, Piscataway, NJ 08855-
0849}
\maketitle
\date{January 30, 1998}
\maketitle
\begin{abstract}
We use inelastic neutron scattering to measure the 
magnetic fluctuations in a single crystal of the heavy fermion alloy
CeCu$_{5.9}$Au$_{0 .1}$ close to the antiferromagnetic quantum critical
point.  The energy(E)-, wavevector(q)- and temperature(T)-dependent spectra 
obey E/T scaling at Q near (1,0,0). The neutron data and earlier bulk
susceptibility are consistent with the form 
$\chi^{-1} \sim f(Q)+(-iE+aT)^{\alpha}$, with an anomalous exponent
$\alpha \approx 0.8 \neq 1$. We confirm the earlier observation of quasi-low
dimensionality and show how both the magnetic
fluctuations and the thermodynamics can be understood in terms
of a quantum Lifshitz point.
\end{abstract}
\vskip 0.1 truein
\pacs{PACS numbers: 71.27.+a, 75.20.Hr, 75.40.Gb }
\vskip2pc]

Quantum phase transitions are zero temperature transitions driven by a
parameter such as pressure, magnetic field, or composition which
regulates the amplitude of quantum fluctuations. An important
realization of the last years has been that many of the most interesting
phenomena in condensed matter physics seem to occur near such
transitions. Notable examples include high-temperature
superconductivity \cite{hitc,science} and heavy fermion
behavior.\cite{cca,hf} 
For high-temperature
(cuprate) superconductivity, the quantum critical hypothesis leads very
naturally to one of the key attributes of the normal state, namely that
the energy scale governing spin and charge fluctuations is the
temperature T itself, a property labelled E/T scaling.\cite{hitc}

\begin{figure}
\epsfxsize=2.4 truein
\centerline{\epsfbox{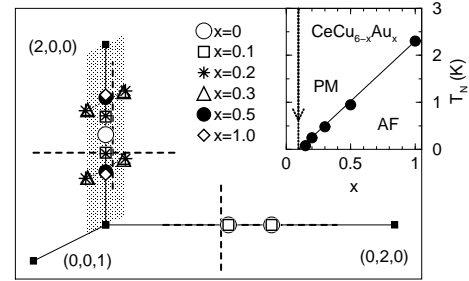}}
\vskip 0.1truein
\caption{
Location of the magnetic Bragg peaks or spin fluctuation maxima  
for CeCu$_{6-x}$Au$_{x}$ at different Au-concentration x 
(from refs \cite{ins,ccanew,0.5,0.2}).
Dotted lines are the trajectories in the (hk0)-plane of the constant-E 
scans presented in Fig.\ \ref{fig2}. Inset shows magnetic ordering 
temperature T$_N$ (from ref \cite{cca}) vs. x. The arrow marks the
thermal trajectory investigated for x=0.1 in this paper.
}
\label{fig1}
\end{figure}

Heavy fermion behavior is generally understood as a consequence
of a competition between intersite spin couplings and the
single-site Kondo effect.\cite{doniach} 
Antiferromagnetic order develops in a heavy fermion metal at a quantum
critical point where the intersite spin couplings 
overcome the disordering Kondo effect of local spin-exchange
processes. While the notion of such competing interactions is
old, the nature of the quantum phase transition 
is a subject of great current interest. 
CeCu$_6$, a heavy fermion metal with one of the largest known
linear specific heats,\cite{cca} provides an ideal opportunity to explore
this physics in quantitative detail.
Here, the
quantum critical point (QCP) occurs as Au is substituted for the Cu atoms;
when more than 0.1 Cu sites per Ce are replaced, the heavy fermion
paramagnet gives way to an ordered antiferromagnet.\cite{cca} 
Fig.\ \ref{fig1} illustrates
the associated phase diagram and ordering vectors. In spite of the large
and rapidly growing literature on this particular QCP
\cite{cca,ccath}, the E/T scaling which has been so thoroughly tested 
for the single crystal cuprates \cite{hitc} and for polycrystalline 
UCu$_{5-x}$Pd$_x$ \cite{U}
has not yet been observed for  single crystal heavy fermion
systems in general and CeCu$_{6-x}$Au$_{x}$ in particular. 
Our purpose here is to
report the discovery, obtained from magnetic neutron scattering,  of E/T
scaling near the QCP in CeCu$_{5.9}$Au$_{0.1}$. Beyond 
establishing
the universality of E/T scaling among QCPs hosted by
vastly different alloys, our data yield a non-mean-field value for a
key critical exponent characterizing the quantum phase transition
in CeCu$_{5.9}$Au$_{0.1}$.

We used the Czochralski technique to grow a single crystal of
CeCu$_{5.9}$Au$_{0.1}$ with volume  $\sim$4 cm$^3$ and a mosaic of
2.5$^\circ$.  To establish the homogeneity and composition of the
sample, we used high-resolution neutron diffraction in the orthorhombic
[001] zone employed for our inelastic neutron scattering measurements.
The result was that throughout 
the whole sample, the room temperature (Pnma) lattice
constants were a = 8.118(2) \AA \, and b = 5.100(1) \AA \, 
corresponding to the Au concentration of x= 0.10$\pm$0.03.\cite{ccax}
Furthermore, the transition into the low-temperature monoclinic phase 
occurred at a location-independent T$_S$ =  70$\pm$2 K, again confirming the
homogeneity of the sample. Between T$_S$ and 5K,  the angle $\gamma$
grows from 90$^o$ to  90.7$^o$, while the a-axis direction remains unchanged.
Thus, the monoclinic distortion is sufficiently small that we follow
past custom and use orthorhombic notation to label
points in reciprocal space.

After installing the sample in the Bell Laboratories/Ris{\o} dilution
refrigerator, we collected inelastic neutron scattering data using 
TAS7 at the Ris{\o} DR3 reactor equipped with a PG(002) 
monochromator and analyzer and a BeO filter.
The final neutron energy was fixed at  E$_f$= 3.7 meV. The
energy resolution,  defined as the full-width-at-half-maximum (FWHM) of the
elastic incoherent scattering from the sample, was 0.134 meV.

\begin{figure}
\epsfxsize=3.2 truein
\epsfbox{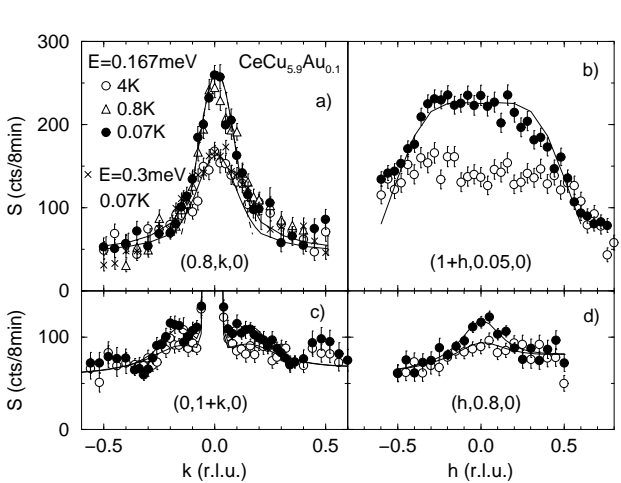}
\caption{
Constant-E scans through the magnetic scattering from
CeCu$_{5.9}$Au$_{0.1}$ at 
E$_0$=0.167 meV near Q=(0.8,0,0) (a,b) and Q=(0 1$\pm$0.15 0) (c,d).
Solid lines in (a,b) correspond to Eq.(3), corrected for experimental 
resolution and with the same parameters as shown in Fig.\ \ref{fig3} 
and f(Q) expanded in even powers in Q. The lines in (c,d) are simply 
guides to the eye. The T-independent peak at (010) in (c) is of
nuclear origin.}
\label{fig2}
\end{figure}

When we began our experiments, we knew that the most intense magnetic
fluctuations in the paramagnetic parent CeCu$_{6}$ were located close to the
inequivalent reciprocal lattice points (010) and (100).\cite{ins} At 
the same time, the
incommensurate vectors (1$\pm\delta$ 0 0) fully describe the
magnetic order in CeCu$_{6-x}$Au$_x$ with x$\ge$0.5 \cite{ccanew,0.5} 
and do so partially for x= 0.2.\cite{0.2}
Thus, to discover the
dominant fluctuations near the QCP at x= 0.1,  we investigated
fluctuations with wave-vectors of both types. Fig.\ \ref{fig2} shows the
corresponding constant-energy scans at E$_0$= 0.167meV, collected
along the trajectories
indicated in Fig.\ \ref{fig1}. Frames (c) and (d) demonstrate that
incommensurate peaks survive at  (0 1$\pm$0.15 0), and that their intensity
grows upon cooling. The widths of the (0 1$\pm$0.15 0) peaks in the (1,0,0) 
and (0,1,0) directions are similar and not resolution-limited, corresponding
to magnetic correlations with a range of order 8\AA (at this
E$_0$). Even though they are still present and growing with decreasing T, 
the (0 1$\pm$0.15 0) peaks
are five times weaker than the scattering, shown in frames (a) and (b),
near (100). Here, there is a weak modulation along (1,0,0) with hints
of maxima at (1$\pm\delta$ 0 0) for $\delta \approx$ 0.2, and a much sharper
modulation along (0,1,0). Both the peak intensity and width along
(0,1,0) are clearly T and E-dependent, corresponding to growing
antiferromagnetic correlations with decreasing T and E. Furthermore, T and E
appear interchangeable in the sense that the profiles 
for E=0.3meV $>$ k$_B$T=0.006meV and E=0.167meV $<$ k$_B$T=0.345meV are 
indistinguishable. That the larger of E and k$_B$T controls the width  
$\kappa$(E,T)
is expected for quantum critical points and has been previously observed in 
the context of
high-temperature superconductors.\cite{science}

\begin{figure}
\epsfxsize=2.7 truein
\centerline{
\epsfbox{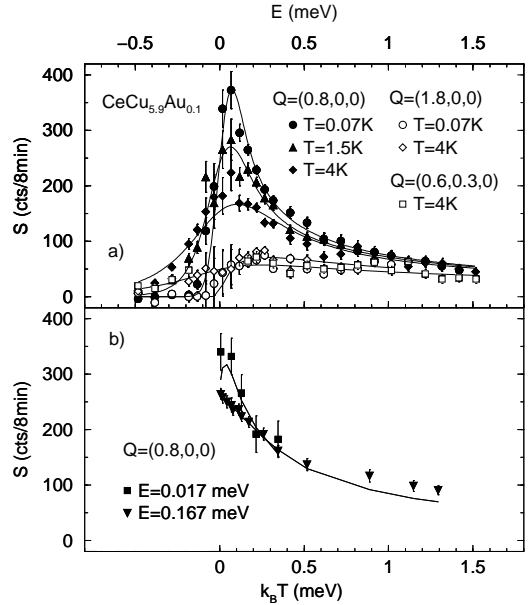}}
\caption{
Constant-Q scans at Q=(0.8 0 0) and Q=(1.8 0 0) and (0.6 0.3 0) at 
different temperatures T
vs energy tranfer E (a) and at different E vs T (b).
Lines correspond to  Eq(3) with c=83cts/8min$\cdot$meV$^{0.74}$, where $a$ and 
$\alpha$ are given by the
analysis of the scaled data (see Fig.\ \ref{fig4} and text) 
and include resolution corrections. 
The optimal (T- and E-independent) values for f(Q) are 
0.036(4)meV$^{0.74}$ and 0.38(2)meV$^{0.74}$ for Q=(0.8 0 0) and the other 
Q-values respectively. 
Note that f(0.8 0 0) is finite, indicating that our
sample is slightly subcritical. But it is sufficiently small
that when the data are examined outside the regime where the resolution matters, 
the deviations from criticality are negligible. 
}
\label{fig3}
\end{figure}

We next characterize the spectrum of the most intense fluctuations,
namely those at Q=(1$\pm$0.2 0 0). Fig.\ \ref{fig3}
shows such constant-Q spectra for several T. 
\noindent For comparison, it also
shows data at  Q=(1.8 0 0) and (0.6 0.3 0), for which no special Q- 
or T-dependent
enhancement is found. The Q- and T-independent elastic scattering
background has been subtracted from all data shown. At Q=(0.8 0 0), the
spectra narrow dramatically with reduction in T, to the point where
they have a rising edge at low E which is comparable to that of an 
instrumentally broadened step function. Thus, as happens near any second 
order phase
transition, cooling results in a reduced magnetic relaxation 
rate.  Even so, a long high-E tail of the spectra persists at all T 
that the FWHM never approaches that of the instrumental
resolution. At Q=(1.8 0 0), the spectra change much less dramatically.

\begin{figure}
\epsfxsize=3.2 truein
\centerline{
\epsfbox{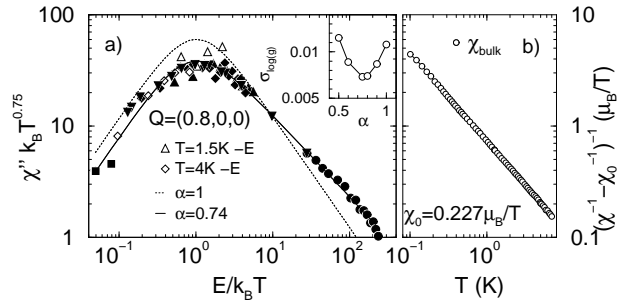}}
\caption{
(a) Scaling plot of the data from Fig.\ \ref{fig3} for Q=(0.8 0 0).
The inset shows how the quality of the scaling collapse varies with
$\alpha$ in Eq.(3), using a procedure unbiased towards any particular form
of the scaling function g(y) (see text). Solid and dashed lines correspond to
Eq.(4) with $\alpha$=0.74 and 1, respectively. 
(b) Double logarithmic plot of susceptibility $\chi_{bulk}$ (from M/B
in $B_c$ =0.1T from \cite{cca}) vs. T after subtracting a zero T value
$\chi_0$ corresponding to a 'parallel' shunt, as suggested by Eq.(3).}
\label{fig4}
\end{figure}

In addition to performing measurements as a function of E at fixed T,
we have also collected data as a function of T for fixed E (Fig.\
\ref{fig3}). The similarity between the T-scans (b) and the E-scans (a) 
suggests  $E/T$ scaling, which we have tested by fitting
the data to the form
\begin{equation}
\chi''(E,T)=  T^{-\alpha} g( E/k_BT).
\end{equation}
No assumptions were made about $g(x)$ beyond the ability to approximate it
by a histogram with stepsize 0.1 on a log$_{10}$ E/T scale 
between -0.5 and 0.7.
The best collapse of the data onto a single curve, 
as checked by the smallest (log) deviation ($\sigma_{log(g)}$ see inset) from 
the
mean step values, is obtained for $\alpha=0.75\pm0.1$ and shown in 
Fig.\ \ref{fig4}, which displays all the data at Q=(0.8 0 0) unaffected by 
the finite experimental resolution.

Having confirmed $E/T$ scaling, we ask whether the associated
exponent $\alpha$ is consistent with other data and theory. The simplest
mean field approach \cite{th}
dictates a susceptibility $\chi$
\begin{equation}
\chi(Q,E,T)^{-1}= c^{-1}(f(Q)-iE+aT),
\end{equation}
where f(Q) is a smooth function of Q with minima at the wave-vectors
characterizing the eventual magnetic order.  If we
position ourselves at the critical Q and composition, f(Q)=0 and
$\chi''= cE/[(aT)^2+E^2]$, which we can
compare to Eq.(1) to read off the scaling function
$g_o(y)= y/(1+y^2)$ with $y= E/aT$ and the exponent $\alpha$=1. The dotted line
in Fig.\ \ref{fig4}(a) corresponds to the (log) optimized $g_o(y)$ with 
a/k$_B$=1 and deviates clearly from the data.

Neither 
$g_o(y)$ nor $\alpha$= 1 account for the measurements; thus
the simple mean-field theory fails for CeCu$_{5.9}$Au$_{0.1}$.
To make progress, we modify Eq(2) to account for arbitrary values of
$\alpha$ via the simple expedient of replacing $(-iE+aT)$ in Eq.(2) by
$(-iE+aT)^{\alpha}$,
\begin{equation}
\chi^{-1}= c^{-1}(f(Q)+(-iE+aT)^{\alpha})
\end{equation}
Beyond accounting for the anomalous value of the exponent of T in the
scaling relation, this generalization has two other important
implications. The first is that the scaling function (for Q at which
f(Q)= 0) becomes
\begin{equation}
g(y)= c \sin[\alpha \tan^{-1}(y)]/(y^2+1)^{\alpha/2}
\end{equation}
The solid line in Fig.\ \ref{fig4} corresponds to the best fit of Eq.(4) to
the data (with c=88cts/8min$\cdot$meV$^{0.74}$); note that the value obtained
from floating
\noindent
$\alpha$ in Eq.(4) in the fitting process is at 0.74$\pm$0.1 indistinguishable
from that obtained as the exponent for the thermal prefactor $\alpha$ in 
Eq.(1).
At the same time, $a/k_B$=0.82 is a number close to
unity, indicating that the magnetic fluctuations have an underlying
energy scale very close to the thermal energy. 

The second consequence of the generalized form (3) is that for $E$=0,
$\chi'(Q,T)^{-1}=c^{-1}(f(Q)+(aT)^{\alpha}$), which implies that
$(\chi'(Q,T)^{-1}-\chi'(Q,0)^{-1})^{-1}= c(aT)^{-\alpha}$.We test this 
relation by using the very accurate and dense data provided by bulk 
magnetometry \cite{cca}
for a special value of Q, namely Q=0. The fact that on the double logarithmic 
scale of Fig.\ \ref{fig4}(b) the data lie on a straight line shows that 
the relation is indeed 
satisfied. The line is not parallel to the diagonal for which
$\alpha$=1, but indeed has slope $\alpha$=0.8$\pm$0.1, indistinguishable from 
$\alpha$ obtained from the neutron scattering data by the 
other two methods described above.

The neutron results are consistent with the bulk
susceptibility: do they also account for the thermodynamics?
One of the key features of this system, is 
that, at low T, $C/T = \gamma(T)$  diverges logarithmically.\cite{cca}
To see whether $\gamma(T) \sim {\rm ln}(T)$
is consistent with the neutron measurements, we need
to consider the detailed dependence of $f(Q)$.
In keeping with earlier investigations \cite{ccath},  the
critical fluctuations are almost two-dimensional. Specifically, 
the constant E scans along  (1,0,0) (Fig.\ \ref{fig2}(b)) look flat-topped,
indicating that the quadratic term in the power series expansion of
$f(Q)$ around
(100) almost vanishes in this direction. A good fit to the data in 
the vicinity of $Q=(0.8 0 0)$ is given by 
$f((100)+\delta Q) =
(0.036+12(\delta k)^2 -0.08 (\delta h)^2 +2 (\delta h)^4)meV^{0.74}$ 
(see dashed/solid lines in Fig.\ \ref{fig2}).
That the quadratic term B$_h(\delta h)^2$ is negative (with a positive 
quartic term
C$_h (\delta h)^4$) also follows from the 
observation that CeCu$_{6-x}$Au$_x$ orders not at (100)
but at pairs of incommensurate wavevectors nearby (see Fig.\ \ref{fig1}).
As the sign of B$_h$ 
changes, the minimum at $\delta Q$ = 0 of f($\delta Q$) splits into the pair 
$\delta Q= \pm\sqrt{-B_h/2C_h}$.
In the classical theory of phase transitions, vanishing of the
quadratic stiffness occurs at a Lifshitz point \cite{lifshitz}.
Here, we are dealing with a quantum Lifshitz point. 

To understand the thermodynamics of a quantum Lifshitz point, 
we assume that the quadratic stiffness vanishes along only one
direction (e.g.(1,0,0)) in reciprocal space or more
generally, that it vanishes for a finite number of lines centered
at certain points Q in reciprocal space \cite{lifshitz}. 
The most important contributions to $\chi$ will then be proportional to 
$1/[B_{\perp} (\delta Q_{\perp})^2 + C_{\parallel} (\delta Q_{\parallel})^4
+(-iE+aT)^{2 \over z}]$ in the vicinity of $Q$, where 
$\delta Q_{\parallel}$ and $\delta Q_{\perp}$
measure the departure from $Q$ parallel and perpendicular to
the line direction, and we have employed the notation of 
dynamical critical phenomena, writing $\alpha = 2/z$.
If $\kappa_{\perp}$ and $\kappa_{\parallel}$
are the inverse correlation lengths in these directions
\[
\kappa_{\perp} \sim T^{1/z}, \qquad \kappa_{\parallel}\sim T^{1/z_{
\parallel}}=T^{1/2z},
\]
so the dynamical critical exponent is $2z$ along the $(1,0,0)$
direction, but $z$ in the perpendicular directions. This has the
effect of reducing the effective dimensionality D$_{eff}$. 
For an isotropic QCP, the free energy scales as
$
F\sim T \kappa^D \sim T^{1 + D/z}.
$
For a quantum Lifshitz point, we must now write
\[
F\sim 
T \kappa_{\perp}^{D-1} \kappa_{\parallel} 
\sim  T^{1 + (D-\frac{1}{2})/z}.
\]
So D$_{eff}$ is lower by {\sl one half}, and 
\[
\gamma(T) = - \partial_T^2 F(T) \sim T^{ (D-\frac{1}{2})\alpha/2-1}.\]
We see that for $D=3$, $\alpha =0.8$, 
the exponent for $\gamma(T)$ vanishes, in accord with the specific
heat data. If we ignore the anisotropic form of f(Q) indicated by the 
neutron scans through reciprocal space and assume a conventional (non-Lifshitz) 
QCP for CeCu$_{6-x}$Au$_x$, the  $\alpha$ obtained from 
the energy- and temperature-dependence of the neutron data as 
well as the bulk $\chi$ would lead to $\gamma\sim T^{0.2}$ and $T^{-0.2}$
for D=3 and 2 respectively, in disagreement with the C(T) results.

Our ability to use the phenomenological
scaling form for $\chi^{-1}$ for both the thermodynamics {\sl and}
the uniform susceptibility indicates that the anomalous
energy dependence of the susceptibility extends over a wide
region of the Brillouin zone.  
We can write the T=0 susceptibility in
(3) as $\chi^{-1}(Q,E)=\chi^{-1}(E)-J(Q)$, where 
$\chi(E)$ describes the response of individual local moments and 
surrounding conduction electrons and $J(Q)$ defines the interactions between
different screened local moments. Eq.(3) actually determines $\chi^{-1}(E)$ 
and $J(Q)$ up to the same additive constant $\Theta$
which sets the scale for the screening:
$\chi^{-1}(E)=c^{-1}(-iE)^{\alpha}+\Theta$ and
$J(Q)=-c^{-1}f(Q)+\Theta$. 
Our result that $\alpha=0.8\pm 0.1$ implies that $\chi(E)$
is nonanalytic near E=0.
In the ``Millis-Hertz'' theory of a
quantum critical point, the soft magnetic fluctuations
couple to a Fermi liquid and the corresponding local response is
analytic, with $\alpha=1$.
The anomalous exponent $\alpha\ne 1$ for CeCu$_{5.9}$Au$_{0.1}$
thus represents a fundamental {\sl local} deviation from Fermi liquid
behavior.
One possible interpretation is that the magnetic
QCP has qualitatively modified the underlying moment compensation 
mechanism,\cite{larkin} causing the local electron propagator to develop an 
anomalous scaling dimension. \cite{anderson}  

We have performed the first detailed single crystal measurements of the
magnetic fluctuation spectrum in a heavy fermion system close to an
antiferromagnetic quantum critical point. We find that 
the data obey E/T scaling with an anomalous scaling exponent $\alpha
\approx 0.8$.
The same exponent appears in three other independent quantities: the scaling 
function itself, previously published bulk
susceptibility,\cite{cca} and the specific heat. To obtain the scaling
relation connecting C(T) with the magnetic measurement, we take into account
the (again independent) indications of our neutron data that the QCP in 
CeCu$_{5.9}$Au$_{0.1}$ is actually near a quantum Lifshitz point, 
whose hallmark is the near degeneracy of ordered states with
characteristic wavevectors along some line(s) in reciprocal
space. The general idea  that there are many potential  ground states for 
CeCu$_{5.9}$Au$_{0.1}$ is further reinforced by our
observation of an enhanced susceptibility at the seemingly  inequivalent
locations (0 1$\pm$0.15 0) and (1$\pm$0.2 0 0) in reciprocal space.

We gratefully acknowledge discussions with G. Kotliar,
G. L. Lonzarich, H. v. L\"ohneysen, A. Rosch and O. Stockert. 
Research at Karlsruhe was supported by the Deutsche Forschungsgemeinschaft.
Research at Rutgers was supported in part by the National Science
Foundation under grant NSF DMR 96-14999.

\end{document}